\documentclass{article}

\usepackage[utf8]{inputenc}
\usepackage[T1]{fontenc}

\PassOptionsToPackage{table, dvipsnames}{xcolor}
\usepackage{xcolor}

\PassOptionsToPackage{colorlinks=true, citecolor=blue, linkcolor=blue, urlcolor=black}{hyperref}

\usepackage{arxiv}
\usepackage[utf8]{inputenc}
\usepackage[T1]{fontenc}
\usepackage{url}
\usepackage{booktabs}
\usepackage{nicefrac}
\usepackage{microtype}
\usepackage{lipsum}
\usepackage{graphicx}
\usepackage{natbib}
\usepackage{doi}
\usepackage{algorithm}
\usepackage{multirow}
\usepackage{multicol}
\usepackage{amsmath,amssymb,amsfonts}
\usepackage{amsthm}
\usepackage{mathrsfs}
\usepackage{float}
\usepackage{algorithmic}
\usepackage{bm}
\usepackage{caption}
\usepackage{hyperref}

\title{Velocity model building from seismic images using a Convolutional Neural Operator}

\author{ \href{https://orcid.org/0009-0009-0709-9158}{\includegraphics[scale=0.06]{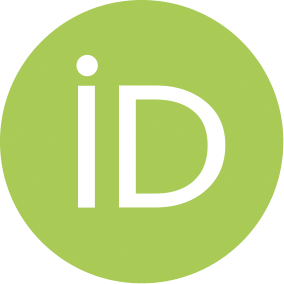}\hspace{1mm}Xiao~Ma}\\
	Division of Physical Science and Engineering\\
	King Abdullah University of Science and Technology\\
	Thuwal 23955-6900, Saudi Arabia \\
	\texttt{xiao.ma@kaust.edu.sa} \\
        \And
	\href{https://orcid.org/0000-0002-9363-9799}{\includegraphics[scale=0.06]{orcid.png}\hspace{1mm}Tariq~Alkhalifah} \\
	Division of Physical Science and Engineering\\
	King Abdullah University of Science and Technology\\
	Thuwal 23955-6900, Saudi Arabia \\
	\texttt{tariq.alkhalifah@kaust.edu.sa} \\
}

\renewcommand{\shorttitle}{Velocity model building using Neural Operator}
\graphicspath{{./Figure/}}
\hypersetup{
pdftitle={A template for the arxiv style},
pdfsubject={q-bio.NC, q-bio.QM},
pdfauthor={David S.~Hippocampus, Elias D.~Striatum},
pdfkeywords={First keyword, Second keyword, More},
}

\begin{document}
\maketitle

\begin{abstract}
The success of building a high-resolution velocity model using machine learning is hampered by generalization limitations that often limit the success of the approach on field data. This is especially true when relying on neural operators for the mapping. Thus, we propose a novel inversion framework that relies on learning to map the velocity model to a seismic image using a Convolutional Neural Operator (CNO), and then we use optimization to invert for the velocity that matches the image. The key to the success of our network is that we use the initial and true velocity models as input in the training, then we invert for the true velocity starting from the initial velocity at inference. Specifically, we first train a neural operator to accurately learn the forward mapping from seismic velocity models to RTM images, using synthetic datasets that include high-frequency structural information. Once trained, the neural operator is embedded into an inversion loop, where its differentiable nature enables efficient gradient computation via automatic differentiation. This allows us to progressively inject high-wavenumber information from RTM images into the background velocity model, thereby improving resolution without the need for traditional adjoint-state solvers. The proposed framework is validated on both synthetic and field data. Results demonstrate that the neural operator generalizes well to real seismic scenarios, maintains high inversion accuracy, and significantly reduces computational cost. This work highlights the potential of neural operators as flexible and scalable tools for efficient, data-driven seismic imaging and inversion.
\end{abstract}

\section{\textbf{Introduction}}
In recent years, deep learning has emerged as a transformative tool in seismic imaging and inversion, offering a powerful alternative to traditional physics-based algorithms. Unlike traditional inversion methods, which mainly depend on physical formulations, deep learning-based inversion frameworks are generally categorized into two types: data-driven inversion methods and unsupervised approaches that do not rely on training data. Data-driven inversion methods leverage the powerful nonlinear representation and the learning capabilities of neural networks. In these approaches, the input to the network typically consists of seismic data—such as shot gathers \citep{yang2019deep}, migrated images, or angle-domain common-image gathers \citep{zhang2022deep}—while the target output corresponds to subsurface properties such as velocity or acoustic impedance \citep{zhao2019stochastic}. The neural network is trained to learn the nonlinear mapping from seismic observations to the underlying geological structure by continuously optimizing its parameters based on the data. Among the various supervised learning approaches in seismic inversion, direct inversion from raw shot gathers to subsurface velocity models has received significant attention. This class of methods directly leverages the nonlinear representation capabilities of neural networks. For instance, \cite{li2019deep} constructed a series of relatively realistic synthetic velocity models to evaluate the performance of neural network–based direct inversion. Their experiments demonstrated that the network was able to accurately recover velocity structures on the test dataset, producing results that closely matched the corresponding ground truth models. Furthermore, \cite{kazei2020velocity} utilized common midpoint gathers as input to predict velocity profiles and applied this method to field seismic data. The results showed that, while the neural network was able to capture the overall structural outline of the subsurface, the finer details in the predicted velocity model remained highly correlated with patterns present in the synthetic training data. This observation highlights a significant limitation of such direct inversion approaches: their generalization capability is often inadequate when transitioning from synthetic to real-world data. To mitigate the limitations associated with direct inversion methods, several studies have explored the use of migrated images as auxiliary inputs to enhance the network's performance. For example, \cite{zhang2020adjoint} incorporated RTM (Reverse Time Migration) images into the inversion framework as a representation of high-wavenumber information. By feeding both the RTM image and the corresponding background velocity model into the network, their approach effectively alleviated the cycle-skipping problem commonly encountered in full waveform inversion (FWI). Building upon this idea, \cite{yang2023well} further integrated well log information with RTM images, aiming to strengthen the physical constraints and improve generalization. Their results demonstrated that such multimodal fusion significantly enhances the network's robustness, yielding accurate inversion results even for out-of-distribution (OOD) test sets. Despite these advances, the inherent dependence of supervised learning methods on large volumes of synthetic data remains a key bottleneck. This data limitation continues to constrain their applicability to real seismic scenarios, especially when the synthetic training sets fail to capture the full complexity of field data. \par

In addition to direct inversion methods, an increasing number of studies have explored unsupervised learning approaches to address the limitations of conventional full waveform inversion, such as dependence on initial velocity models and the issue of cycle skipping \citep{yang2023fwigan}. Within this framework, neural networks primarily function as regularizers. Compared to traditional regularization techniques, such as total variation, neural network-based regularization can be directly embedded into each stage of the inversion process without the need to re-derive gradients.  In particular, \cite{saad2025enhancing} proposed a novel framework based on a Siamese neural network architecture to enhance the resolution of full-waveform inversion. In this framework, the observed and simulated seismic data are fed into two identical subnetworks that share the same weights. By comparing the latent feature representations extracted from both inputs, the network computes a similarity-based loss that guides the inversion process. Numerical experiments demonstrate that incorporating the Siamese network significantly improves the resolution of the inverted models, particularly in complex geological settings. Unsupervised inversion methods exhibit application potential due to their reduced reliance on labeled data and enhanced generalization capabilities \citep{wang2024controllable}. However, several challenges remain to be addressed. In particular, the training process often involves high computational costs, primarily due to the need for iterative forward and backward modeling. Additionally, the optimization of such networks—especially when incorporating complex architectures or physics-based constraints—can be difficult and sensitive to hyperparameter settings \citep{ma2025effective}. Overcoming these limitations will be critical for enabling the practical deployment of unsupervised neural inversion frameworks in large-scale, real-world geophysical applications.\par

As for the neural network architecture used in the deep learning based inversion frameworks, most data-driven inversion methods are built upon convolutional neural networks (CNNs). CNNs have demonstrated remarkable success in computer vision and related fields, primarily due to the advantages of convolution operations in capturing spatial correlations and hierarchical features in image-like data. These characteristics make CNNs particularly suitable for geophysical inversion tasks, where seismic data can often be interpreted in image form.  Currently, a variety of neural network architectures have gained popularity in the context of seismic inversion, including models based on VGG-16 \citep{simonyan2014very,feng2021multiscale}, ResNet \citep{he2016deep, wu2022seismic}. In addition to CNNs, a relatively novel class of models—neural operators—has been attracting increasing attention in both the machine learning and geophysical communities. Neural operators are specifically designed to learn mappings between infinite-dimensional function spaces, rather than finite-dimensional vectors. This makes them particularly well-suited for problems involving partial differential equations (PDEs) and continuous field representations. Due to their ability to generalize across varying spatial resolutions and domain geometries, neural operators have shown promising performance in complex geophysical applications, such as forward modeling \citep{li2023solving, lehmann2023fourier, kong2023feasibility, huang2025learned} and direct velocity inversion that mapping shot gather into velocity models \citep{molinaro2023neural, zhu2023fourier}, where traditional neural networks may struggle to scale or generalize effectively. Given the high accuracy demonstrated by neural operators in solving forward problems \citep{yang2021seismic, li2023solving, song2022high}, recent research has begun to exploit the differentiable nature of neural operators to perform velocity model inversion using automatic differentiation. For instance, \cite{zou2024deep} proposed a method in which a neural operator is trained to learn the mapping from a three-dimensional velocity model to the corresponding frequency-domain wavefield. Once the network is trained, gradients of the wavefield with respect to the velocity model can be efficiently computed via the chain rule, enabling velocity updates. Experimental results demonstrate that neural operators accurately capture the frequency-domain wavefield responses and enable effective gradient-based inversion.\par

In this study, we propose a novel neural operator–based inversion framework that explicitly bridges Reverse Time Migration (RTM) imaging and velocity model reconstruction. The proposed method leverages a convolutional neural operator to learn the mapping from velocity models—including both background velocity and true velocity—to RTM images. The forward operator is trained using synthetic data enriched with high-frequency components to ensure that the network can capture fine-scale structural features. Once trained, the operator remains fixed and is embedded into an inversion loop, where automatic differentiation is used to update the background velocity model such that the predicted RTM image aligns with the one obtained from observed data. This framework is tested on both synthetic and field seismic datasets. Our results demonstrate that the proposed method not only generalizes well from synthetic to real data, but also effectively propagates high-frequency information from the RTM image into the inverted velocity model. Moreover, the approach achieves significant computational acceleration by avoiding repeated numerical solutions of wave equations, making it a compelling candidate for large-scale seismic inversion tasks.

\section{Theory}
In this section, we start by describing the training dataset involved in training the forward machine learning network that maps velocities to the corresponding RTM image. The neural operator used for the mapping is then described. We finally describe the framework for using this trained network to invert for a high-resolution velocity model. 
\subsection{Training dataset}
\begin{figure}[htbp]
\centering
\includegraphics[width=\textwidth]{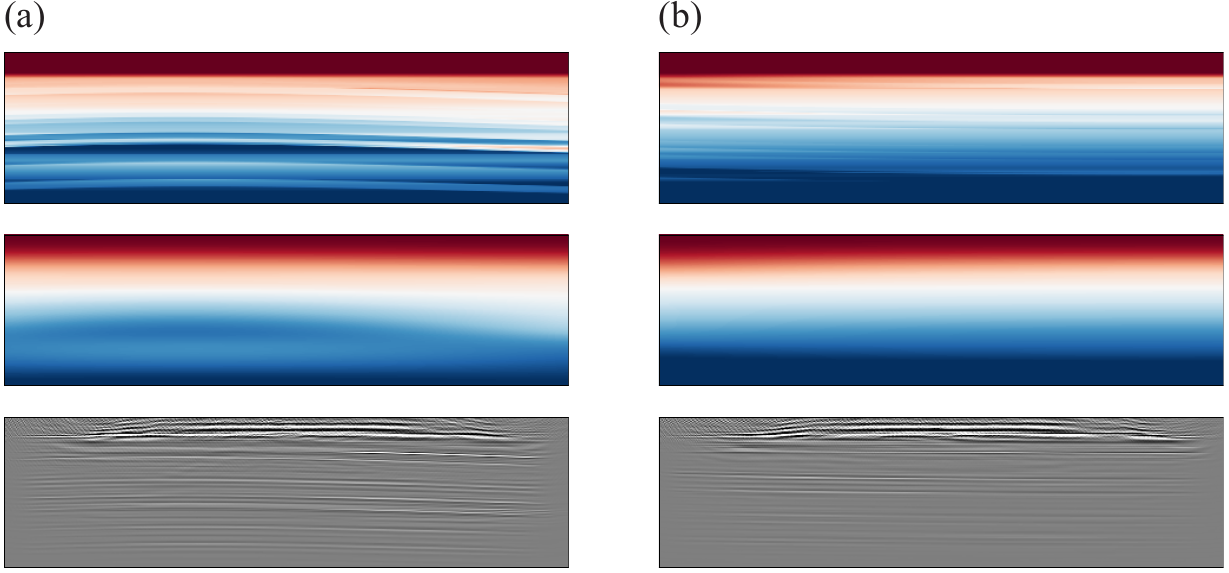}
\caption{ The two columns (a and b) represent two sampled velocity models used during neural network training. From top to bottom, each row corresponds to the true velocity model, the background velocity model, and the associated RTM image, respectively.}
\label{fig1} 
\end{figure}

To enable the network to effectively learn, in a supervised fashion, the relationship between velocity models and the corresponding migrated images, we propose using synthetic seismic velocity models as training samples. The advantage of employing synthetic data lies in the ability to generate a large number of training samples. By varying the structural characteristics of these models, the network is able to learn a broader range of prior information, thereby improving its generalization capability. Figure~\ref{fig1} shows two instances of the training samples. The training dataset consists of layered velocity models with varying layer thicknesses, and the velocity range differs from one layer to another. For each velocity model, we use a smoothed version of the corresponding true velocity model as the background velocity, which is used to obtain the corresponding migrated image from the simulated data using the reverse time migration method.

\subsection{Operator learning}
In this chapter, we employ a neural operator to address an operator learning problem.  The neural operators are a class of models composed of linear integral operators and non-linear activations.  These neural operators incorporate a nonlinear point-wise activation function following each linear integral operator. A neural operator with $L$ layers is formulated in an iterative manner as follows:
\begin{equation}
v_0(x) = P (a(x)),
\end{equation}
\begin{equation}
v_{l+1}(x) = \sigma\left( W_l v_l(x) + \int_D \kappa_l(x, y)\, v_l(y)\, dy \right), \quad l = 0, \dots, L - 1,
\end{equation}
\begin{equation}
u(x) = Q (v_L(x)),
\end{equation}
here, $a(x)$ represents the input function(s), such as the synthetic velocity model, while $u(x)$ is the output function(s) (e.g., corresponding RTM image). The variable $v_l(x)$ is the hidden state at the $l$-th layer and serves as the input to the subsequent layer. The operator $P$ is a point-wise lifting function that maps the input into a higher-dimensional latent space, where $Q$ is a projection operator that reduces the final hidden representation back to the target output space. The term $W_l$ is a learnable point-wise linear operator designed to capture non-periodic boundary behaviors and also functions as a residual connection. The kernel $\kappa_l(x, y)$ is a learnable parametric kernel, and $\sigma$ denotes a point-wise nonlinear activation function such as Rectified linear unit (ReLU) or Gaussian Error Linear Unit (GELU).

\begin{figure}[htbp]
\centering
\includegraphics[width=\textwidth]{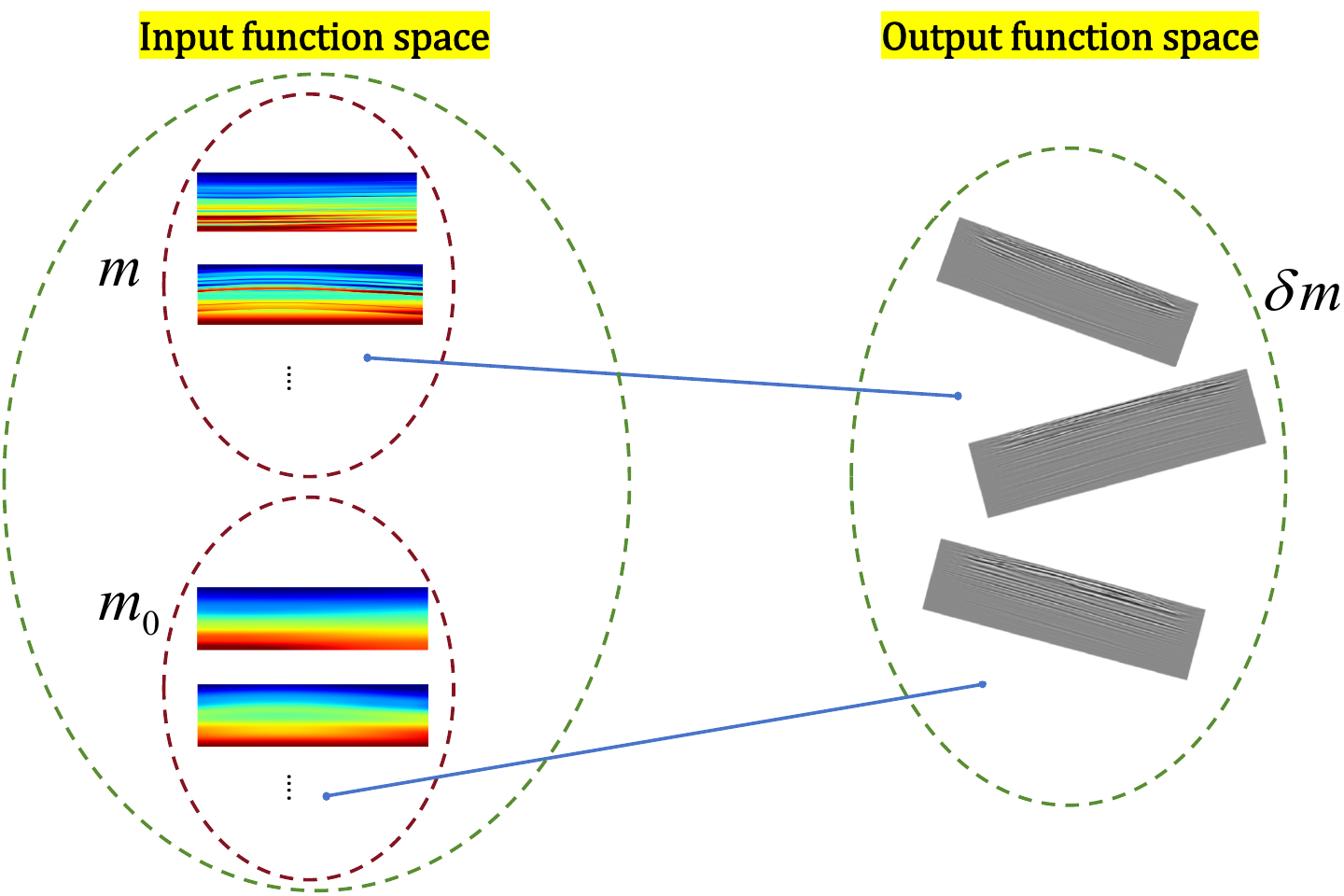}
\caption{In this operator learning framework, the input function space is composed of both the true velocity model and the background velocity model. The target function space to be learned is the migrated image, which predominantly contains high-frequency information.}
\label{fig2} 
\end{figure}

In the context of our operator learning problem, the background velocity model primarily contains low-frequency information, while the RTM image encodes high-frequency structural details. Based on this observation, we aim to train a neural operator to learn the mapping from low-frequency to high-frequency representations. Specifically, as illustrated in Figure~\ref{fig2}, during the training phase, the input to the neural operator consists of the true velocity model—which contains the full range of information—and the background velocity model, which retains only the low-frequency components. The RTM image, which captures predominantly high-frequency features, serves as the target function in the output function space.

We introduce a neural operator architecture designed to learn the mapping from seismic velocity models to corresponding RTM images. The network adopts a U-Net-like encoder-decoder structure, in which each block is constructed based on the ResNet-101 architecture. This design leverages the deep residual learning capability of ResNet-101 to improve feature extraction and convergence stability during training.

\begin{figure}[htbp]
\centering
\includegraphics[width=\textwidth]{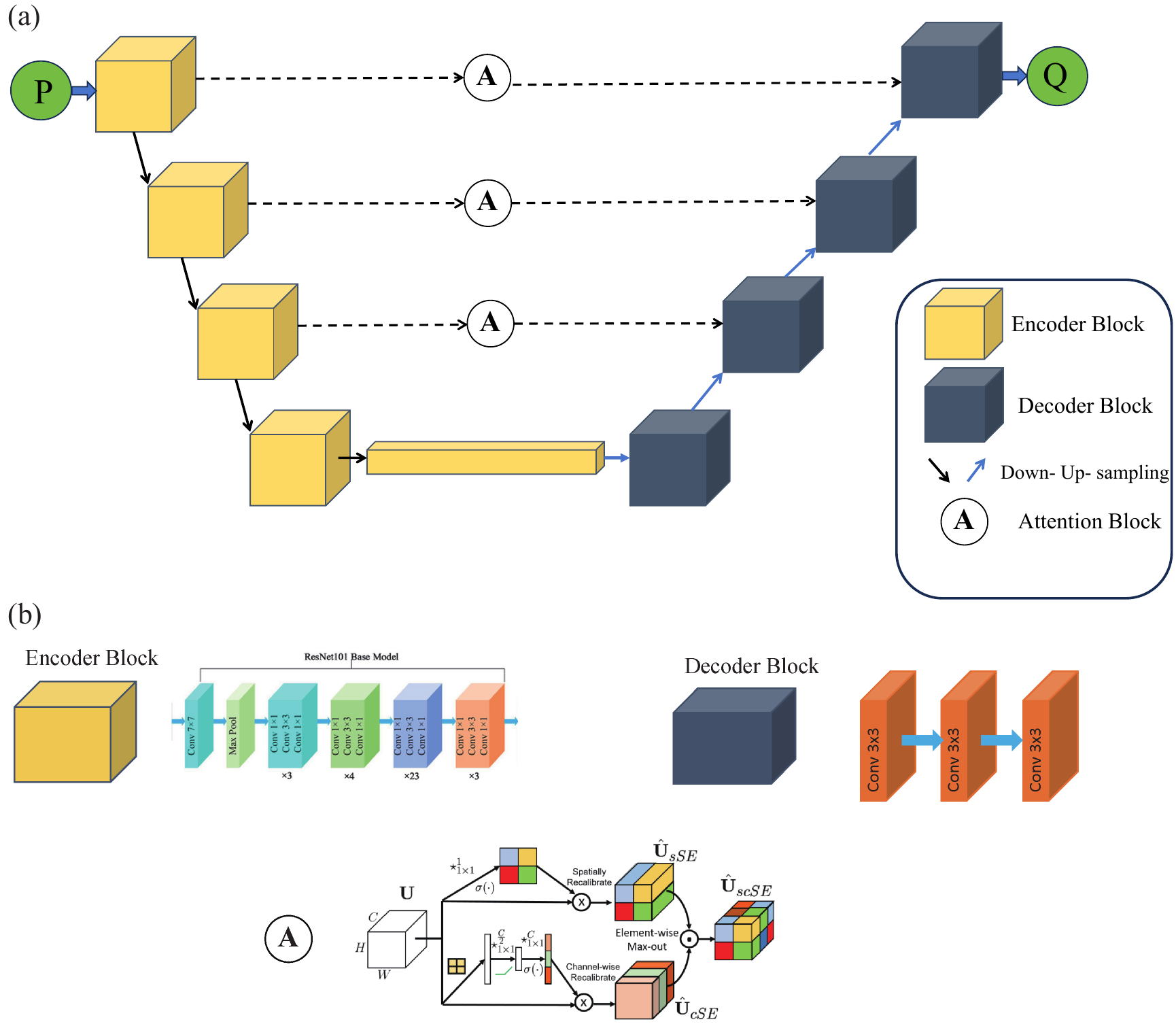}
\caption{The detailed architecture of the proposed neural operator. The components $P$ and $Q$ represent the lifting and projection operators, respectively, which map the input and output between the physical and latent function spaces. Each encoder block is constructed based on the ResNet-101 backbone, each decoder block comprises a sequence of three convolutional layers, designed to progressively reconstruct the target representation from the encoded features. Furthermore, an attention mechanism is incorporated at the skip connection stage to enhance feature fusion between corresponding encoder and decoder layers. (b) provides a detailed view of the internal structure of each encoder and decoder block, highlighting the flow of data and the operations involved.}
\label{fig3} 
\end{figure}

As shown in Figure~\ref{fig3}, the encoder consists of multiple stages, each comprising several residual blocks derived from ResNet-101. These blocks include the identity and convolutional shortcuts to preserve gradient flow and enable the learning of deep hierarchical representations.

The decoder mirrors the encoder structure and is composed of multiple decoding blocks. Each decoder block contains a sequence of three convolutional layers, each with a kernel size of $3 \times 3$, followed by batch normalization and ReLU activation. These layers progressively reconstruct high-resolution spatial features from the compressed latent representations.

To enhance the effectiveness of skip connections between encoder and decoder, we employ two types of attention mechanisms \citep{chen2017sca}:
\begin{itemize}
    \item \textbf{Channel Attention (CA):} This module recalibrates feature responses by modeling inter-channel dependencies. It is implemented via global average pooling followed by a two-layer fully connected network and a sigmoid activation.
    \item \textbf{Spatial Attention (SA):} This mechanism emphasizes salient spatial regions by applying a $7 \times 7$ convolution over the concatenated average-pooled and max-pooled feature maps along the channel axis.
\end{itemize}

These attention modules are integrated into each skip connection to guide the decoder in selecting informative features, thus improving the quality of the predicted RTM image.

As formulated in equation~\eqref{eq:NO}, the input to the network consists of the background velocity model $v$ and the corresponding true velocity $v_{true}$, while the output is the RTM image $\mathcal{I}_{\text{RTM}}$:
\begin{equation}
\mathcal{I}_{\text{RTM}} = \mathcal{F}_{\theta}(v, v_{true})
\label{eq:NO}
\end{equation}
where $\mathcal{F}_{\theta}$ denotes the neural operator with trainable parameters $\theta$.

\subsection{Velocity model building using an image}

\begin{figure}[htbp]
\centering
\includegraphics[width=\textwidth]{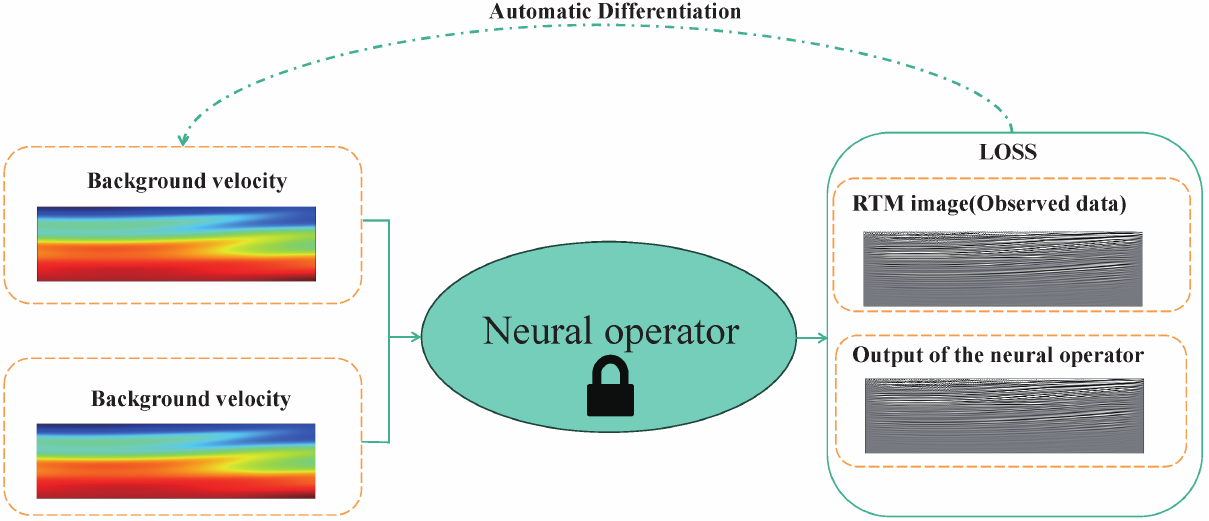}
\caption{During the inversion phase, the pretrained neural operator is kept fixed, and the channel used to input the true velocity model in the training stage is replaced by the background velocity model. The inversion is driven by minimizing the difference between the output of the neural operator and the observed seismic data (i.e., the RTM image corresponding to the current background velocity). The gradient of the loss function with respect to the input velocity is computed automatically via automatic differentiation. This gradient is then used to iteratively update the current velocity model, guiding it toward a structure that better matches the observed RTM image.}
\label{fig4} 
\end{figure}

Once the neural operator has been fully trained, it is frozen and we adopt an inversion-based strategy to iteratively update the velocity model. Figure~\ref{fig4} shows the inversion framework; it is important to note that the gradient of the velocity model with respect to the seismic data (i.e., the RTM image) is automatically computed via automatic differentiation. More specifically, during the inversion phase, both input channels of the neural operator are initially set to the same background velocity model. We then iteratively update the channel used to input the true velocity model in the training stage by minimizing the discrepancy between the neural operator’s output and the observed RTM image. This process is repeated until the predicted RTM image closely matches the actual one, indicating convergence of the velocity model toward a plausible solution. Through this inversion framework, I aim to embed the high-frequency information contained in the RTM image into the background velocity model via automatic differentiation. By leveraging the gradient flow from the RTM-based misfit, the inversion process guides the low-frequency background velocity to gradually incorporate high-frequency structural details, thereby improving the resolution of the inverted velocity model.

\section{\textbf{Numerical examples}}
In this section, we present results obtained using the proposed neural operator-based inversion framework. During the experiments, we observed that this framework exhibits high computational efficiency and can be readily applied to real seismic data with minimal modification. These findings suggest that the method has strong potential for practical deployment in seismic imaging and inversion tasks.

\subsection{training details}
\begin{figure}[htbp]
\centering
\includegraphics[width=0.8\textwidth]{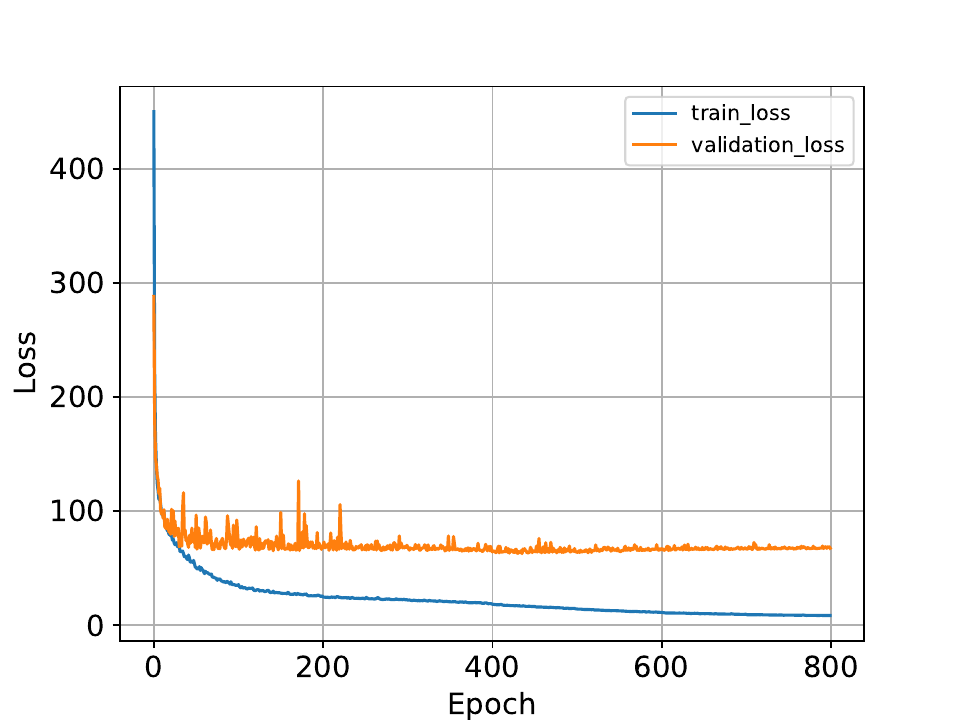}
\caption{Training and validation losses during the training phase.}
\label{fig5} 
\end{figure}

During the training phase, a total of 8,000 synthetic seismic samples are used as the training set, and an additional 2,000 samples are reserved for validation. Samples of the training set are shared in Figure~\ref{fig2} and Figure~\ref{fig4}. The network is trained using the mean squared error (MSE) as the objective function over 800 epochs to ensure adequate convergence of the model parameters. The Adam optimizer is employed for parameter updates, and the entire training process takes approximately 21 hours. It is important to note that the training is performed only once; after training, the inference stage is highly efficient and suitable for rapid application to unseen data. In Figure~\ref{fig5}, the training loss and validation loss both decreased. The trained model is now frozen and will be used to invert synthetic and field data.

\subsection{Synthetic data test}

\begin{figure}[htbp]
\centering
\includegraphics[width=0.9\textwidth]{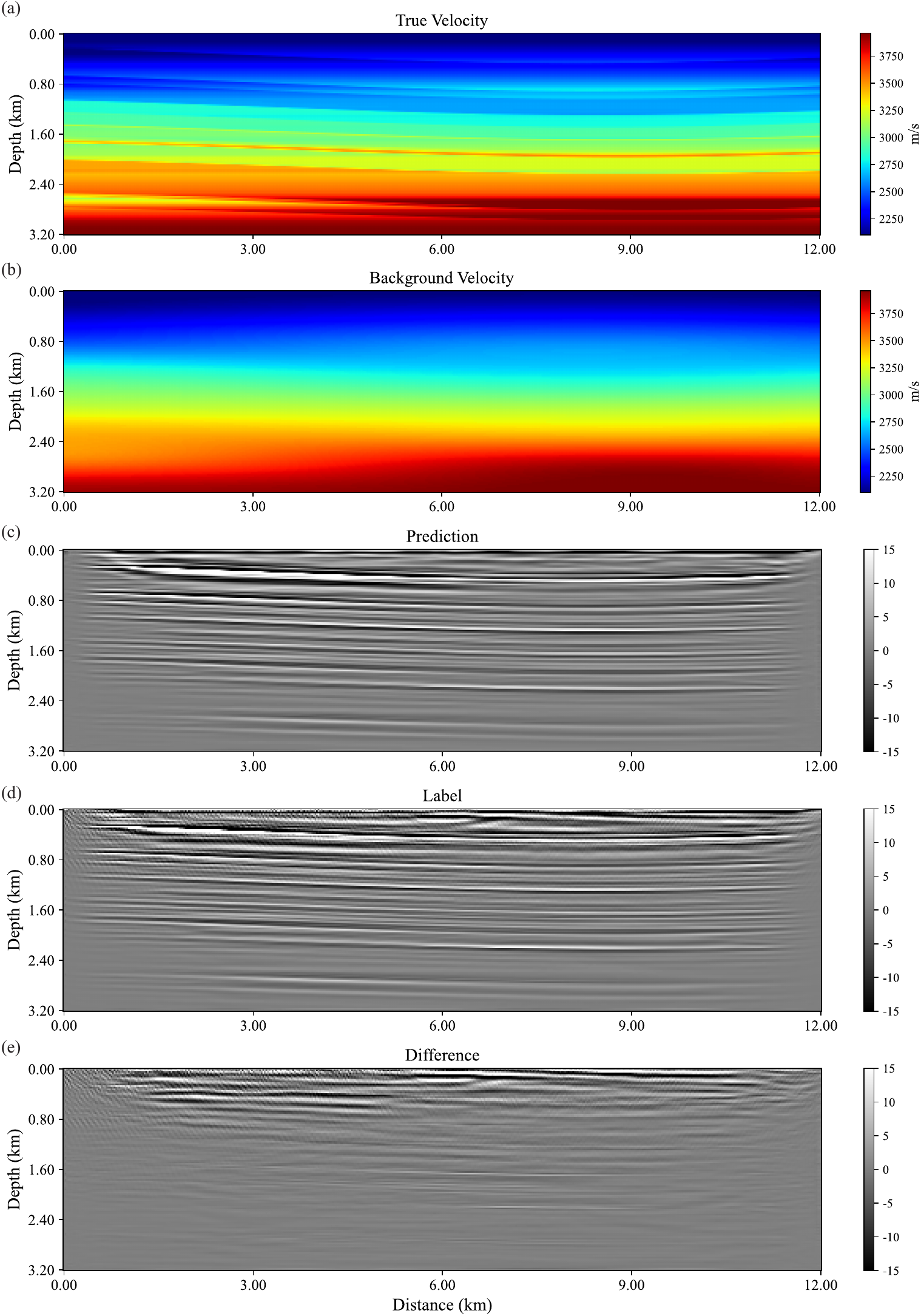}
\caption{Panels (a) and (b) show the ground-truth velocity model and the corresponding background velocity model used in the test example. These two models are concatenated and provided as two input channels to the neural operator. (c) presents the RTM image predicted by the trained network, (d) displays the corresponding ground-truth RTM image used as the label, and (e) shows the difference between the ground-truth and predicted RTM image. The close visual agreement between the predicted and true RTM images demonstrates the neural operator's ability to learn the mapping from the velocity models to seismic images in a forward modeling context.}
\label{fig6} 
\end{figure}

After training the neural operator on synthetic seismic data, we first evaluate its ability to generate seismic images corresponding to velocity models—namely, to perform a forward test. In this test, the input velocity models are taken from a separate test dataset that is not seen during either the training or validation phases. This evaluation serves to assess the generalization capability of the neural operator in predicting RTM images from previously unseen velocity models (in distribution). Figure~\ref{fig6}(a) and Figure~\ref{fig6}(b) show the true velocity model and the corresponding background velocity, respectively, while Figures~\ref{fig6}(c) and ~\ref{fig6}(d) present the predicted RTM image from the neural operator and the ground-truth RTM image. By comparing the predicted and reference images (Figure~\ref{fig6}(e)), it is evident that the neural operator successfully learns the mapping from the velocity model to the migrated image during training, note that the mismatches observed at the top may be attributed to the presence of significant migration-related noise in the shallow layers of the training data. Overall, this result lays a solid foundation for the feasibility of the subsequent inversion process.

\begin{figure}[htbp]
\centering
\includegraphics[width=0.9\textwidth]{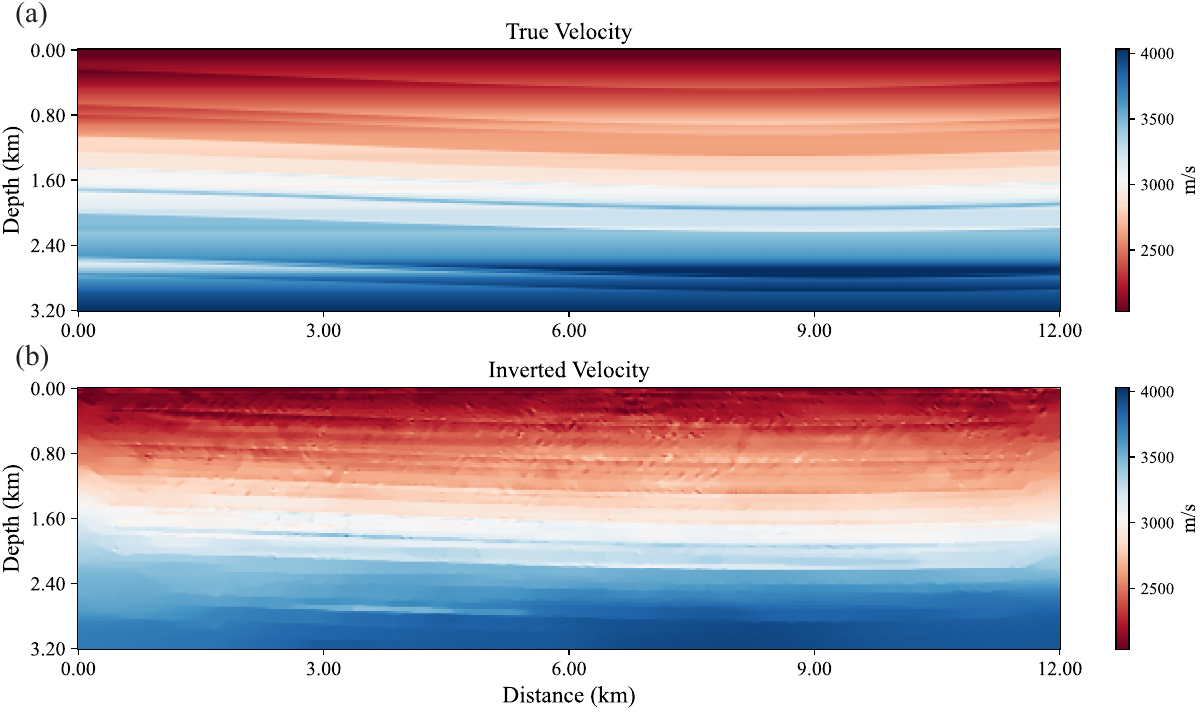}
\caption{(a) shows the true velocity model, (b) presents the velocity model inverted by the neural operator.}
\label{fig7} 
\end{figure}

\begin{figure}[htbp]
\centering
\includegraphics[width=0.5\textwidth]{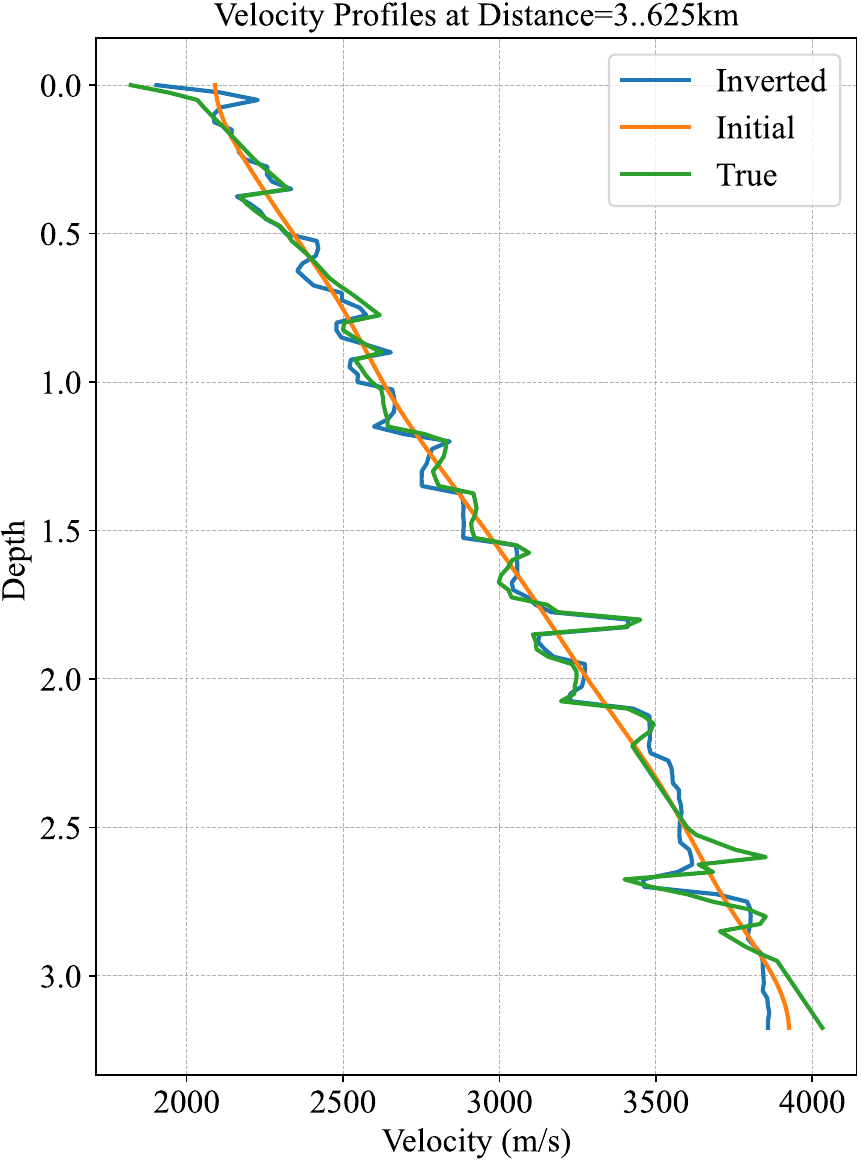}
\caption{Vertical profile (like a well) Comparison in Figure~\ref{fig7}}
\label{fig8} 
\end{figure}

Next, we use the neural operator to invert the high-frequency components of the velocity model and embed them into the background velocity. As described in the "Method" section, the neural operator remains fixed during the inversion phase. Both input channels of the operator are set to the background velocity model, and gradients are computed via automatic differentiation to iteratively update the velocity. During inversion, we update the background velocity over 150 iterations using the Adam optimizer with a learning rate of 15. To ensure stability throughout the inversion process, the objective function combines the mean squared error (MSE) loss with a total variation (TV) regularization term. The inversion result is shown in Figure~\ref{fig7}. Note that it only takes 15 seconds to invert for a single velocity model using a single NVIDIA A100 GPU. By comparing the inversion results with the true velocity model, it can be observed that the inversion method is indeed capable of embedding the high-frequency components present in the RTM image into the background velocity. This is further demonstrated by the well log comparison(Figure~\ref{fig8}), where the inverted well log (blue curve) captures significantly more high-frequency information than the background velocity model (yellow curve).

\begin{figure}[htbp]
\centering
\includegraphics[width=\textwidth]{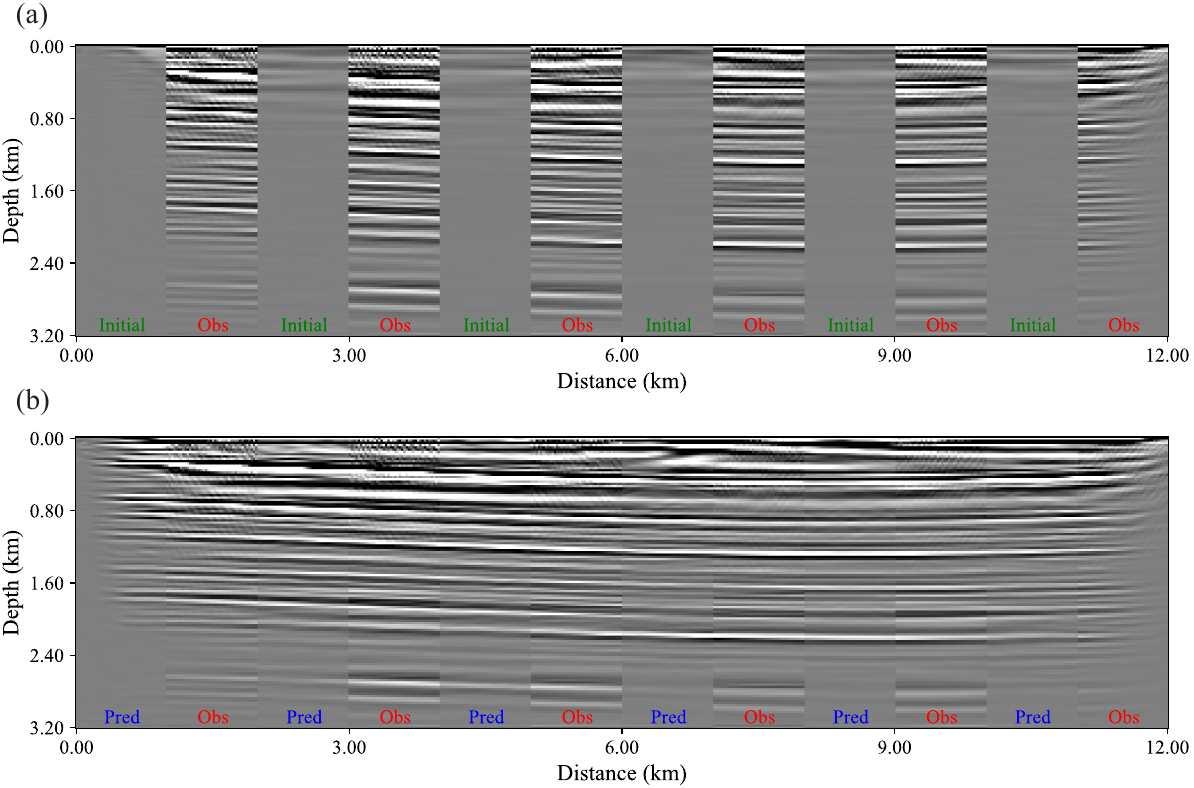}
\caption{Interleaved plots between the target RTM image from synthetic data (observed) and a) the predicted from the initial velocity, b) the predicted from the inverted velocity.}
\label{fig9} 
\end{figure}

Furthermore, the accuracy of data fitting is evaluated to assess the effectiveness of the inversion framework. As shown in Figure~\ref{fig9}, “Initial” refers to the forward modeling result obtained by inputting the background velocity into the neural operator, where most of the values are close to zero. In contrast, “Pred” represents the forward modeling result corresponding to the inverted velocity. One can observe that most of the observed seismic data are well reproduced by the synthetic data generated through the proposed method, indicating a high level of waveform consistency. Owing to the inherent noise-suppression capability of neural networks, much of the shallow noise visible in the RTM image is effectively removed. However, noticeable mismatches appear in regions below a depth of 2.4 km. This result aligns with common geophysical understanding, as deeper subsurface areas typically exhibit lower signal-to-noise ratios and reduced illumination, which collectively limit the accuracy of the inversion in these zones.

\subsection{Field data test}
\begin{figure}[htbp]
\centering
\includegraphics[width=0.9\textwidth]{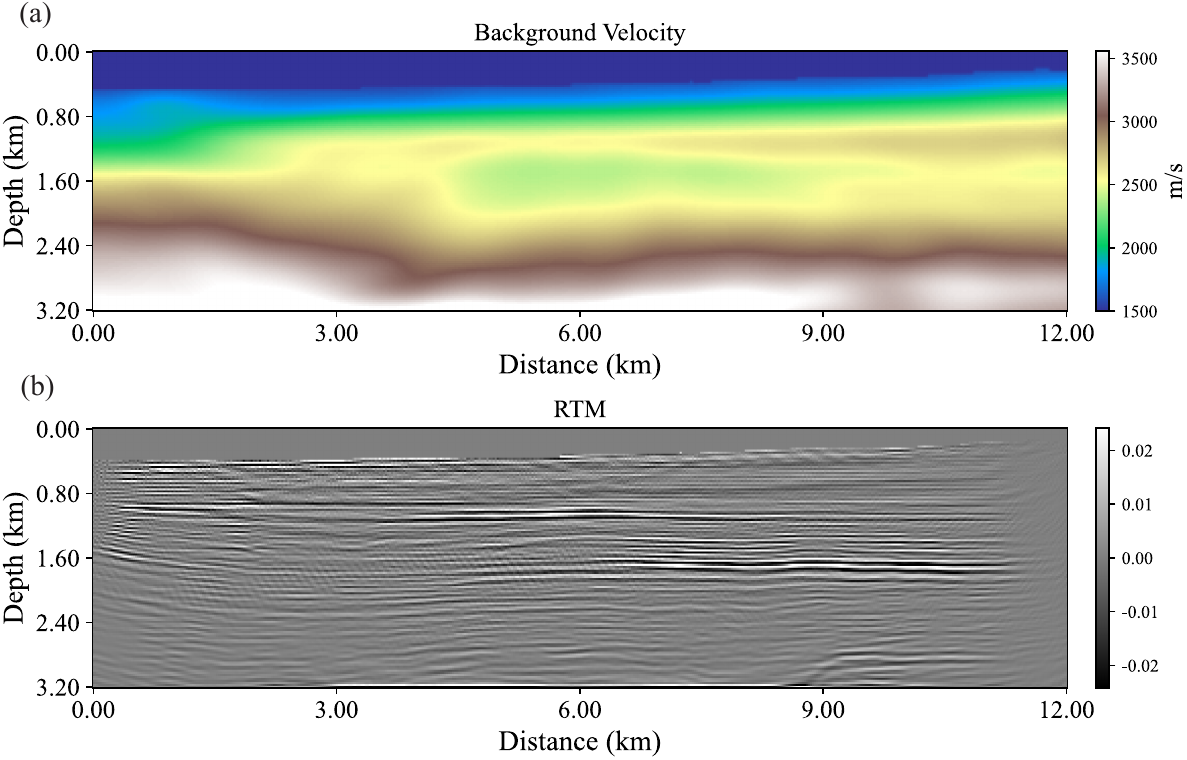}
\caption{Background Velocity Model and Corresponding RTM Image for field Data}
\label{fig10} 
\end{figure}

In this section, we demonstrate the effectiveness of the proposed method by applying it to field data. The neural operator used here is identical to the one employed in the synthetic experiments, which is trained on synthetic velocity models and applied directly to the field data. Specifically, we test the approach on a 2D marine streamer dataset acquired by CGG from the North-Western Australian Continental Shelf. To reduce the computational burden of the inversion, we downsample the original 1824 shots (with a spacing of 18.75 m) to 116 shots. The receiver array consists of 648 channels spaced at 12.5 m, resulting in offsets ranging from 16.9 m to 8256 m. The seismic data were recorded over a maximum duration of approximately 7 seconds, with a temporal sampling interval of 2 ms. The velocity model spans 12 km in the horizontal direction and 3.3 km in depth, discretized on a uniform grid with 12.5 m spacing in both the horizontal and vertical directions \citep{kalita2019flux}. Figure~\ref{fig10} presents the background velocity model obtained through migration velocity analysis, along with the corresponding RTM image. The background velocity appears relatively smooth, containing predominantly low-frequency components with limited high-frequency details. During the inversion stage, this background velocity model derived from field data serves as the initial model, while the RTM image is treated as the seismic data input. \par
\begin{figure}[htbp]
\centering
\includegraphics[width=0.9\textwidth]{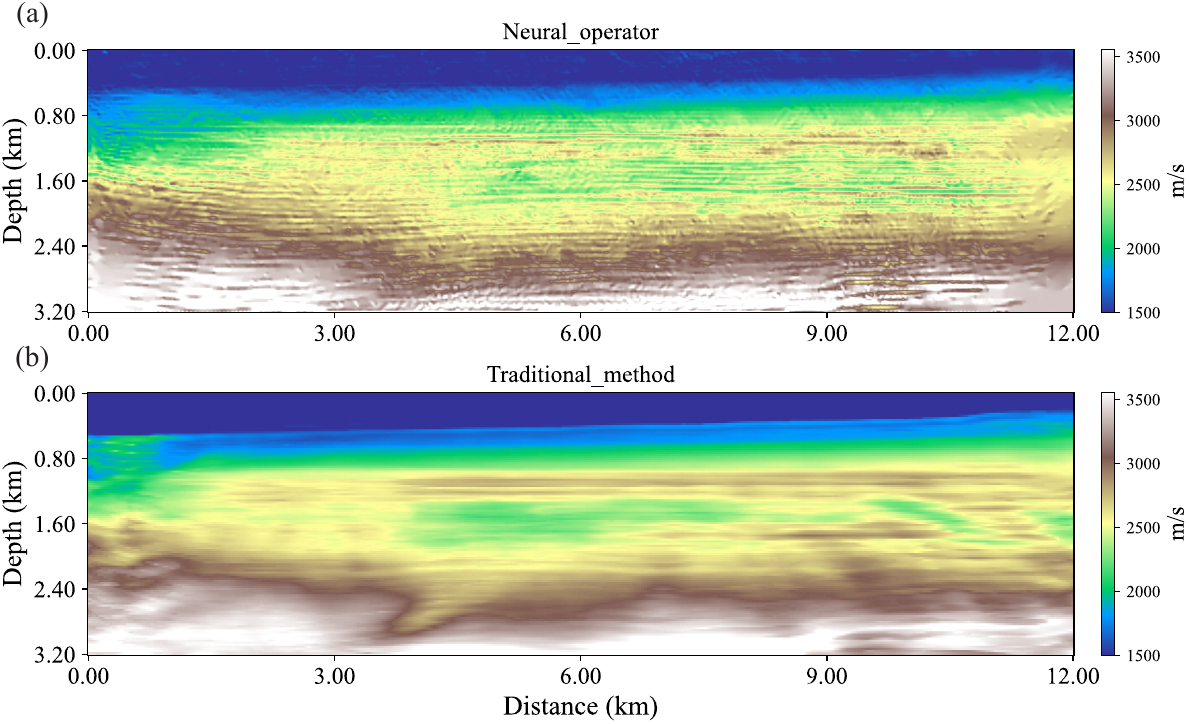}
\caption{Comparison between neural operator-based full waveform inversion and traditional full waveform inversion }
\label{fig11} 
\end{figure}

In Figure~\ref{fig11}, we show the inverted velocity model produced by the neural operator and the traditional method. Specifically, for comparison, we also implement a conventional multi-scale full waveform inversion strategy using frequency bands ranging from 3 to 16 Hz. The inversion is carried out using the Adam optimizer to iteratively update the velocity model, and the objective function is defined based on the cross-correlation between observed and simulated waveforms, which enhances robustness against phase shifts and noise. As shown in Figure~\ref{fig11}(a), the final inverted velocity model is obtained after approximately 2.5 hours of computation. In contrast, the proposed neural-operator-based method achieves comparable or even higher resolution results in a very short time (15s), highlighting its substantial advantage in computational efficiency. Although the network is trained solely on synthetic seismic data, the inverted velocity model exhibits high resolution, clearly revealing high-wavenumber components that are typically difficult to recover. This demonstrates the strong generalization capability of the proposed framework when applied to field data. Moreover, compared to conventional full waveform inversion methods, the proposed approach significantly improves computational efficiency by avoiding repeated numerical solutions of the wave equation during the inversion process. This efficiency improvement makes the method particularly attractive for large-scale inversion applications. \par

\begin{figure}[htbp]
\centering
\includegraphics[width=0.95\textwidth]{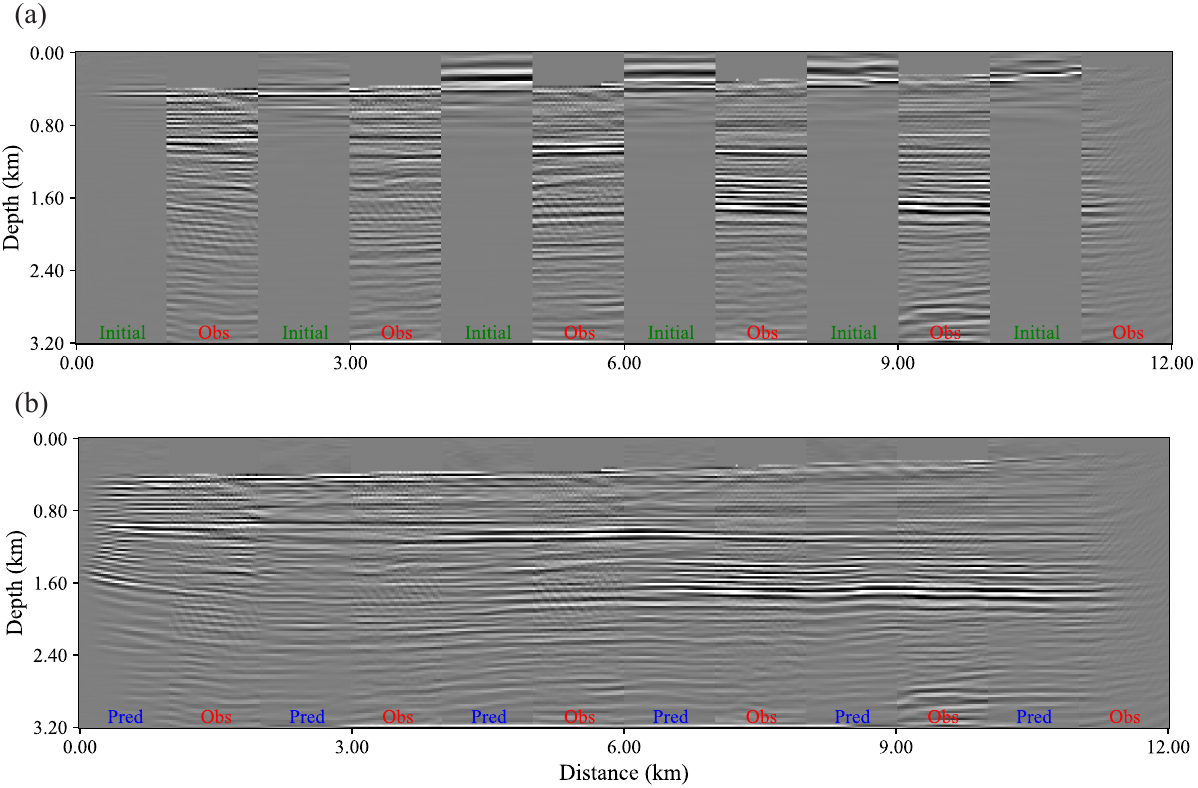}
\caption{Interleaved plots between the target RTM image from field data (observed) and a) the predicted from the initial velocity, b) the predicted from the inverted velocity.}
\label{fig12} 
\end{figure}

In addition, we perform the data comparison. In Figure~\ref{fig12}, when compared with the observed data (i.e., the RTM image), the RTM image generated by the neural operator closely matches the reference (label) RTM image. This agreement indirectly supports the reliability of the inverted velocity model, as it suggests that the neural operator accurately reproduces the high-wavenumber component expected from the RTM imaging. Nevertheless, it is important to notice the limitations of the approach. As a data-driven framework, the performance of the neural network depends strongly on the distribution and representativeness of the training dataset. The generalization ability on field data is closely tied to how well the synthetic training data reflect the geological and geophysical conditions of the target area. Furthermore, because the method relies on RTM images as input, it does not account for complex wave phenomena such as multiples or mode conversions. This limitation may affect inversion accuracy in geologically complex environments, such as areas with strong impedance contrasts or salt bodies. Finally, the method is also dependent on the accuracy of the background velocity, as it is used in imaging as well as an input to the network.

\begin{figure}[htbp]
\centering
\includegraphics[width=0.9\textwidth]{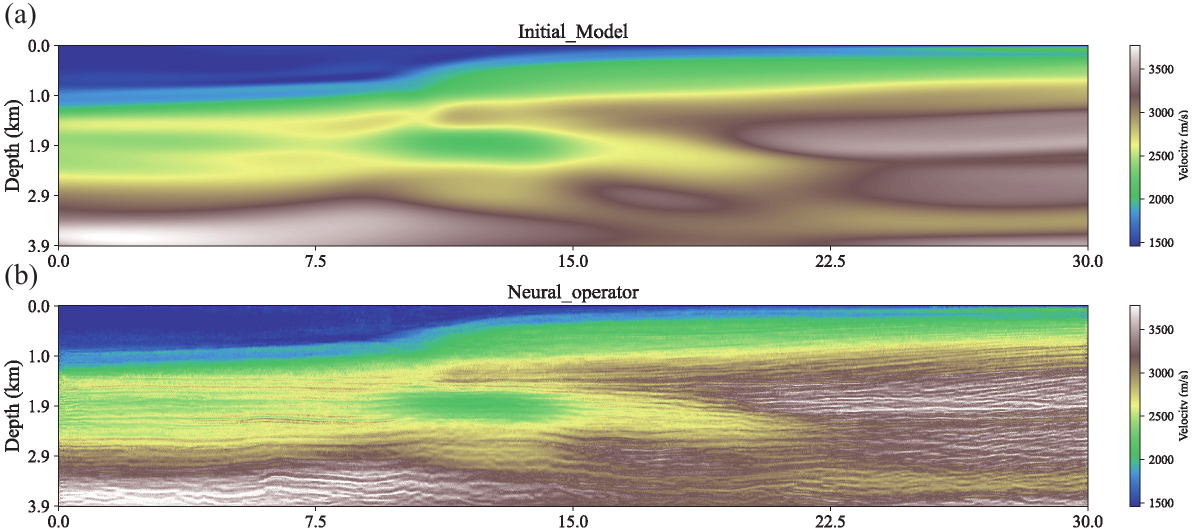}
\caption{A generalization test using a larger-scale field data to evaluate the mesh-independent property of the neural operator. (a) shows the initial background velocity model, (b) presents the velocity model reconstructed by the neural operator.}
\label{fig13} 
\end{figure}

\begin{figure}[htbp]
\centering
\includegraphics[width=0.9\textwidth]{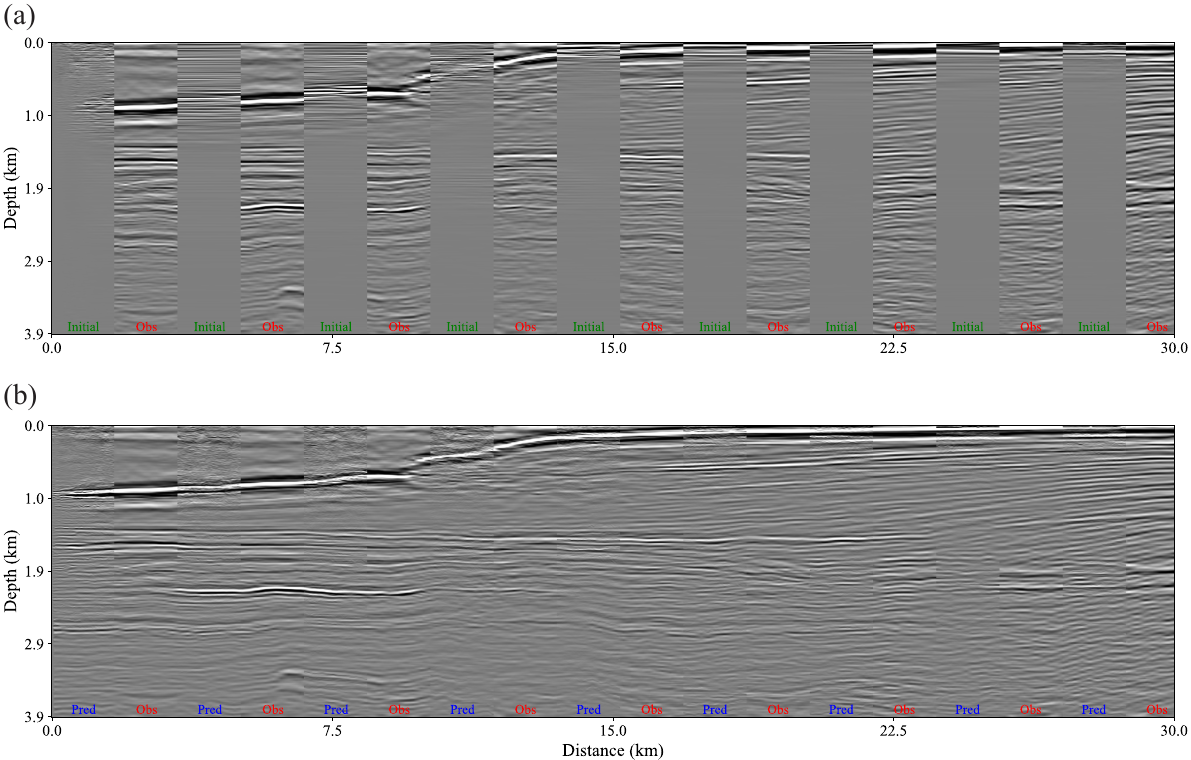}
\caption{Interleaved plots between the target RTM image from field data (observed) and a) the predicted from the initial velocity, b) the predicted from the inverted velocity.}
\label{fig14} 
\end{figure}

Moreover, due to the mesh-independent nature of neural operators, the model is trained on data generated from a fixed spatial grid but can be directly applied to domains with finer or more complex discretizations without the need for retraining. This property significantly enhances the flexibility and scalability of the proposed method, allowing it to adapt seamlessly to higher-resolution scenarios and different mesh configurations while maintaining inversion accuracy. Figure~\ref{fig13} presents results from CGG field data with a significantly larger spatial coverage. The inversion process for this dataset takes approximately 44 seconds to complete, again, demonstrating the computational efficiency of the proposed method. It can be observed that the neural operator is still capable of reconstructing a velocity model with relatively high spatial resolution. Notably, although the model is trained solely on a limited number of synthetic samples defined on relatively small computational grids, it successfully generalizes to the full CGG field data, which spans a length of approximately 30 km. In this case, the high-frequency components embedded in the RTM image are effectively inverted and embedded into the background velocity model (Figure~\ref{fig13}(b)). Furthermore, from the perspective of data fitting, the predicted seismic data (Figure~\ref{fig14}(b)) closely reproduce most of the subsurface geological structures compare with the initial status (Figure~\ref{fig14}(a)), indicating that the inverted model preserves key geophysical features and maintains strong consistency with the observed imaging.

\section{\textbf{Discussion}}

\subsection{Gradient of the inversion stage}
\begin{figure}[htbp]
\centering
\includegraphics[width=\textwidth]{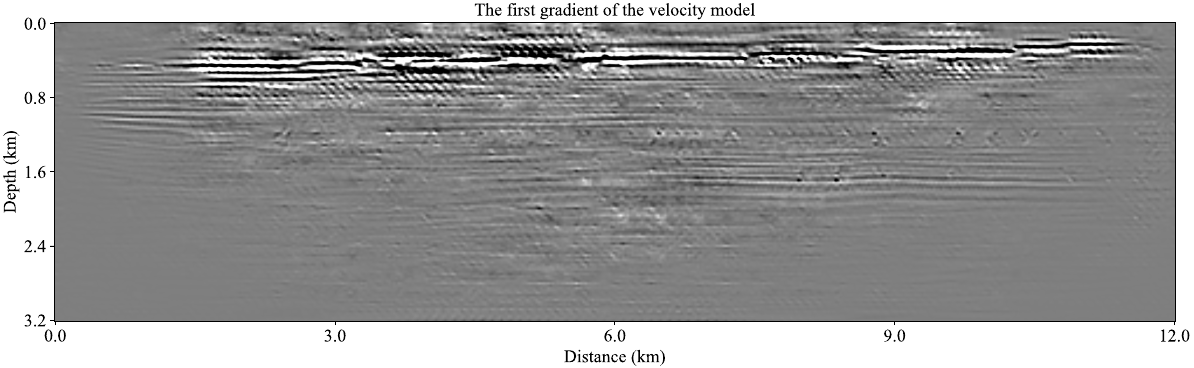}
\caption{The gradient computed during the first update of the velocity model is shown. It can be observed that high-frequency components have been effectively extracted from the RTM image and are clearly reflected in the gradient.}
\label{fig15} 
\end{figure}

Since this study proposes the first attempt to progressively inject high-frequency components from RTM images into the background velocity model using automatic differentiation, it is crucial to analyze the gradient information throughout the inversion process. Gradient analysis provides valuable insights into the underlying optimization dynamics, revealing how different regions of the model are being updated and how effectively the high-wavenumber details are being incorporated. In Figure~\ref{fig15}, one can observe that the gradients computed via automatic differentiation indeed contain high-frequency components present in the observed seismic data. This observation confirms that high-wavenumber information, which is essential for improving resolution, can be effectively extracted from RTM images and propagated into the background velocity model through the gradient flow. However, it is also important to note that the computed gradients exhibit substantial noise. The origin of such noise remains under further investigation.

\subsection{The influence of the initial model}

\begin{figure}[htbp]
\centering
\includegraphics[width=0.9\textwidth]{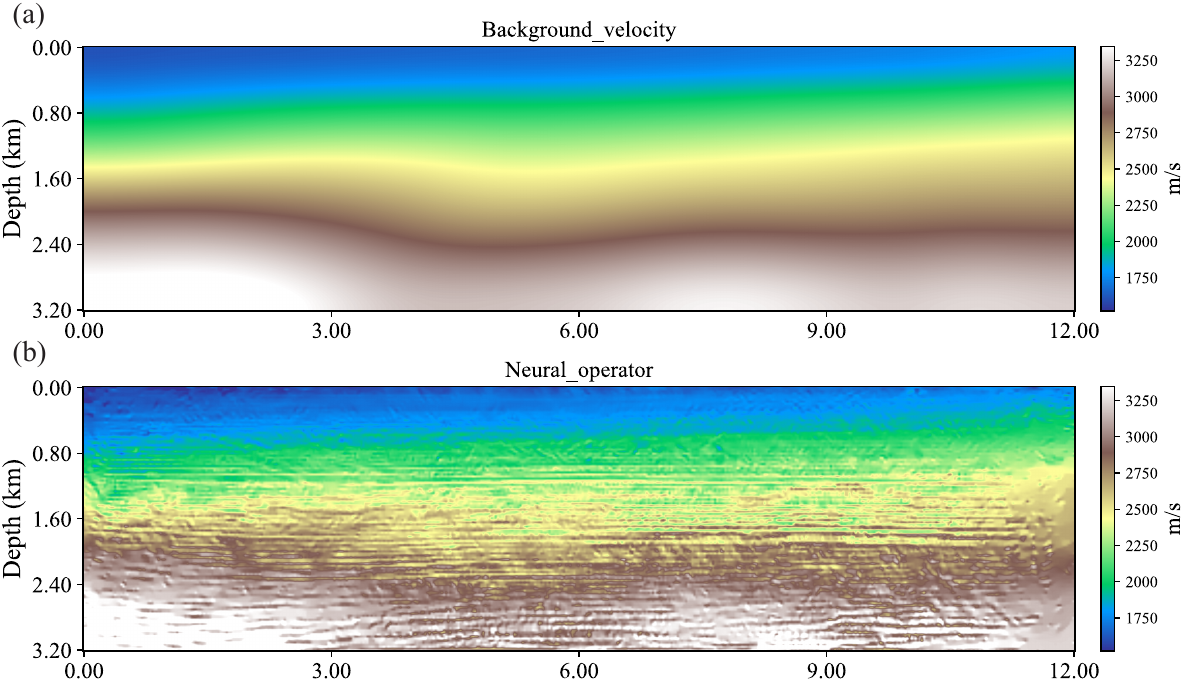}
\caption{(a) shows a very smooth (inaccurate) velocity model, (b) is the inversion result using the neural operator}
\label{fig16} 
\end{figure}

The proposed method relies on RTM images, and therefore, the accuracy of the initial velocity model has a considerable impact on the inversion results. As shown in Figure~\ref{fig16}, we conduct an inversion using a highly smoothed initial velocity model together with its corresponding RTM image. Figure~\ref{fig16}(b) presents the final inverted velocity. Compared with the results obtained using the conventional method (Figure~\ref{fig11}(b)) and our previous inversion Figure~\ref{fig11}(a), we can observe that the accuracy of the shallow velocity is still in need of improvement. This highlights the sensitivity of the method to the quality of the initial model. To rapidly obtain a high-resolution velocity model using this method, careful initial velocity model building is likely required.

\section{\textbf{Conclusions}}
We proposed a neural operator framework that learns the forward modeling task of mapping velocity models to RTM images. The training data are carefully designed to include high-frequency components, enabling the neural operator to capture fine-scale structural features that are essential for high-resolution imaging. Once trained, the network accurately predicts RTM images from velocity models, highlighting its potential for modeling complex wavefield behavior. We further validate the proposed method on field seismic data, where the neural operator is integrated into an inversion workflow. Notably, the high-frequency components originally present in the RTM images are successfully propagated into the inverted velocity model, resulting in improved resolution and structural detail. In addition, the differentiable nature of the neural operator allows for efficient gradient-based inversion using automatic differentiation, achieving a significant reduction in computational cost compared to conventional PDE-based methods. These results confirm the practical applicability of the proposed approach for large-scale seismic imaging tasks, particularly in FWI settings where both accuracy and computational efficiency are critical.\\

\section{\textbf{Acknowledgment}}
This publication is based on work supported by the King Abdullah University of Science and Technology (KAUST). The authors thank the DeepWave sponsors for their support. This work utilized the resources of the Supercomputing Laboratory at King Abdullah University of Science and Technology (KAUST) in Thuwal, Saudi Arabia.
\vspace{0.5cm}
\section{\textbf{Code Availability}}
The data and accompanying codes that support the findings of this study are available at: 
\url{https://github.com/DeepWave-KAUST/Neural_Operator_VMB}. (During the review process, the repository is private. Once the manuscript is accepted, we will make it public.)

\bibliographystyle{unsrtnat}
\bibliography{references}

\begin{thebibliography}{26}
\providecommand{\natexlab}[1]{#1}
\providecommand{\url}[1]{\texttt{#1}}
\expandafter\ifx\csname urlstyle\endcsname\relax
  \providecommand{\doi}[1]{doi: #1}\else
  \providecommand{\doi}{doi: \begingroup \urlstyle{rm}\Url}\fi

\bibitem[Yang and Ma(2019)]{yang2019deep}
Fangshu Yang and Jianwei Ma.
\newblock Deep-learning inversion: A next-generation seismic velocity model building method.
\newblock \emph{Geophysics}, 84\penalty0 (4):\penalty0 R583--R599, 2019.

\bibitem[Zhang et~al.(2022)Zhang, Sun, Zhang, and Zhao]{zhang2022deep}
Jian Zhang, Hui Sun, Gan Zhang, and Xiaoyan Zhao.
\newblock Deep learning seismic inversion based on prestack waveform datasets.
\newblock \emph{IEEE Transactions on Geoscience and Remote Sensing}, 60:\penalty0 1--11, 2022.

\bibitem[ZHAO et~al.(2019)ZHAO, LIU, FENG, GUO, and RUAN]{zhao2019stochastic}
PengFei ZHAO, Cai LIU, Xuan FENG, ZhiQi GUO, and QingFeng RUAN.
\newblock Stochastic seismic inversion based on neural network.
\newblock \emph{Chinese Journal of Geophysics}, 62\penalty0 (3):\penalty0 1172--1180, 2019.

\bibitem[Li et~al.(2019)Li, Liu, Ren, Chen, Yang, Wang, and Jiang]{li2019deep}
Shucai Li, Bin Liu, Yuxiao Ren, Yangkang Chen, Senlin Yang, Yunhai Wang, and Peng Jiang.
\newblock Deep-learning inversion of seismic data.
\newblock \emph{arXiv preprint arXiv:1901.07733}, 2019.

\bibitem[Kazei et~al.(2020)Kazei, Ovcharenko, and Alkhalifah]{kazei2020velocity}
Vladimir Kazei, Oleg Ovcharenko, and Tariq Alkhalifah.
\newblock Velocity model building by deep learning: From general synthetics to field data application.
\newblock In \emph{SEG Technical Program Expanded Abstracts 2020}, pages 1561--1565. Society of Exploration Geophysicists, 2020.

\bibitem[Zhang et~al.(2020)Zhang, Gao, Gao, and Chen]{zhang2020adjoint}
Wei Zhang, Jinghuai Gao, Zhaoqi Gao, and Hongling Chen.
\newblock Adjoint-driven deep-learning seismic full-waveform inversion.
\newblock \emph{IEEE Transactions on Geoscience and Remote Sensing}, 59\penalty0 (10):\penalty0 8913--8932, 2020.

\bibitem[Yang et~al.(2023)Yang, Alkhalifah, Ren, Liu, Li, and Jiang]{yang2023well}
Senlin Yang, Tariq Alkhalifah, Yuxiao Ren, Bin Liu, Yuanyuan Li, and Peng Jiang.
\newblock Well-log information-assisted high-resolution waveform inversion based on deep learning.
\newblock \emph{IEEE Geoscience and Remote Sensing Letters}, 20:\penalty0 1--5, 2023.

\bibitem[Yang and Ma(2023)]{yang2023fwigan}
Fangshu Yang and Jianwei Ma.
\newblock Fwigan: Full-waveform inversion via a physics-informed generative adversarial network.
\newblock \emph{Journal of Geophysical Research: Solid Earth}, 128\penalty0 (4):\penalty0 e2022JB025493, 2023.

\bibitem[Saad and Alkhalifah(2025)]{saad2025enhancing}
Omar~M Saad and Tariq Alkhalifah.
\newblock Enhancing multiparameter elastic full-waveform inversion with a siamese network.
\newblock \emph{The Leading Edge}, 44\penalty0 (5):\penalty0 416a1--416a10, 2025.

\bibitem[Wang et~al.(2024)Wang, Huang, and Alkhalifah]{wang2024controllable}
Fu~Wang, Xinquan Huang, and Tariq Alkhalifah.
\newblock Controllable seismic velocity synthesis using generative diffusion models.
\newblock \emph{Journal of Geophysical Research: Machine Learning and Computation}, 1\penalty0 (3):\penalty0 e2024JH000153, 2024.

\bibitem[Ma and Alkhalifah(2025)]{ma2025effective}
Xiao Ma and Tariq Alkhalifah.
\newblock An effective physics-informed neural operator framework for predicting wavefields.
\newblock \emph{arXiv preprint arXiv:2507.16431}, 2025.

\bibitem[Simonyan and Zisserman(2014)]{simonyan2014very}
Karen Simonyan and Andrew Zisserman.
\newblock Very deep convolutional networks for large-scale image recognition.
\newblock \emph{arXiv preprint arXiv:1409.1556}, 2014.

\bibitem[Feng et~al.(2021)Feng, Lin, and Wohlberg]{feng2021multiscale}
Shihang Feng, Youzuo Lin, and Brendt Wohlberg.
\newblock Multiscale data-driven seismic full-waveform inversion with field data study.
\newblock \emph{IEEE transactions on geoscience and remote sensing}, 60:\penalty0 1--14, 2021.

\bibitem[He et~al.(2016)He, Zhang, Ren, and Sun]{he2016deep}
Kaiming He, Xiangyu Zhang, Shaoqing Ren, and Jian Sun.
\newblock Deep residual learning for image recognition.
\newblock In \emph{Proceedings of the IEEE conference on computer vision and pattern recognition}, pages 770--778, 2016.

\bibitem[Wu et~al.(2022)Wu, Xie, and Wu]{wu2022seismic}
Bangyu Wu, Qiao Xie, and Baohai Wu.
\newblock Seismic impedance inversion based on residual attention network.
\newblock \emph{IEEE Transactions on Geoscience and Remote Sensing}, 60:\penalty0 1--17, 2022.

\bibitem[Li et~al.(2023)Li, Wang, Feng, Yang, and Lin]{li2023solving}
Bian Li, Hanchen Wang, Shihang Feng, Xiu Yang, and Youzuo Lin.
\newblock Solving seismic wave equations on variable velocity models with fourier neural operator.
\newblock \emph{IEEE Transactions on Geoscience and Remote Sensing}, 61:\penalty0 1--18, 2023.

\bibitem[Lehmann et~al.(2023)Lehmann, Gatti, Bertin, and Clouteau]{lehmann2023fourier}
Fanny Lehmann, Filippo Gatti, Micha{\"e}l Bertin, and Didier Clouteau.
\newblock Fourier neural operator surrogate model to predict 3d seismic waves propagation.
\newblock \emph{arXiv preprint arXiv:2304.10242}, 2023.

\bibitem[Kong and Rodgers(2023)]{kong2023feasibility}
Qingkai Kong and A~Rodgers.
\newblock Feasibility of using fourier neural operators for 3d elastic seismic simulations.
\newblock Technical report, Lawrence Livermore National Laboratory (LLNL), Livermore, CA (United States), 2023.

\bibitem[Huang and Alkhalifah(2025)]{huang2025learned}
Xinquan Huang and Tariq Alkhalifah.
\newblock Learned frequency-domain scattered wavefield solutions using neural operators.
\newblock \emph{Geophysical Journal International}, 241\penalty0 (3):\penalty0 1467--1478, 2025.

\bibitem[Molinaro et~al.(2023)Molinaro, Yang, Engquist, and Mishra]{molinaro2023neural}
Roberto Molinaro, Yunan Yang, Bj{\"o}rn Engquist, and Siddhartha Mishra.
\newblock Neural inverse operators for solving pde inverse problems.
\newblock \emph{arXiv preprint arXiv:2301.11167}, 2023.

\bibitem[Zhu et~al.(2023)Zhu, Feng, Lin, and Lu]{zhu2023fourier}
Min Zhu, Shihang Feng, Youzuo Lin, and Lu~Lu.
\newblock Fourier-deeponet: Fourier-enhanced deep operator networks for full waveform inversion with improved accuracy, generalizability, and robustness.
\newblock \emph{Computer Methods in Applied Mechanics and Engineering}, 416:\penalty0 116300, 2023.

\bibitem[Yang et~al.(2021)Yang, Gao, Castellanos, Ross, Azizzadenesheli, and Clayton]{yang2021seismic}
Yan Yang, Angela~F Gao, Jorge~C Castellanos, Zachary~E Ross, Kamyar Azizzadenesheli, and Robert~W Clayton.
\newblock Seismic wave propagation and inversion with neural operators.
\newblock \emph{The Seismic Record}, 1\penalty0 (3):\penalty0 126--134, 2021.

\bibitem[Song and Wang(2022)]{song2022high}
Chao Song and Yanghua Wang.
\newblock High-frequency wavefield extrapolation using the fourier neural operator.
\newblock \emph{Journal of Geophysics and Engineering}, 19\penalty0 (2):\penalty0 269--282, 2022.

\bibitem[Zou et~al.(2024)Zou, Azizzadenesheli, Ross, and Clayton]{zou2024deep}
Caifeng Zou, Kamyar Azizzadenesheli, Zachary~E Ross, and Robert~W Clayton.
\newblock Deep neural helmholtz operators for 3-d elastic wave propagation and inversion.
\newblock \emph{Geophysical Journal International}, 239\penalty0 (3):\penalty0 1469--1484, 2024.

\bibitem[Chen et~al.(2017)Chen, Zhang, Xiao, Nie, Shao, Liu, and Chua]{chen2017sca}
Long Chen, Hanwang Zhang, Jun Xiao, Liqiang Nie, Jian Shao, Wei Liu, and Tat-Seng Chua.
\newblock Sca-cnn: Spatial and channel-wise attention in convolutional networks for image captioning.
\newblock In \emph{Proceedings of the IEEE conference on computer vision and pattern recognition}, pages 5659--5667, 2017.

\bibitem[Kalita and Alkhalifah(2019)]{kalita2019flux}
Mahesh Kalita and Tariq Alkhalifah.
\newblock Flux-corrected transport for full-waveform inversion.
\newblock \emph{Geophysical Journal International}, 217\penalty0 (3):\penalty0 2147--2164, 2019.

\end{thebibliography}

\end{document}